\documentclass[aps,prl,preprint]{revtex4-1}

\begin{document}
\title{Gravitational wave memory in de Sitter spacetime}
\author{Lydia Bieri}
\email{lbieri@umich.edu}
\affiliation{Dept. of Mathematics, University of Michigan, Ann Arbor, MI 48109-1120, USA}
\author{David Garfinkle}
\email{garfinkl@oakland.edu}
\affiliation{Dept. of Physics, Oakland University, Rochester, MI 48309, USA}
\affiliation{Michigan Center for Theoretical Physics, Randall Laboratory of Physics, University of Michigan, Ann Arbor, MI 48109-1120, USA}
\author{Shing-Tung Yau}
\email{yau@math.harvard.edu}
\affiliation{Dept. of Mathematics, Harvard University, Cambridge, MA 02138 USA}

\date{\today}

\begin{abstract}
We examine gravitational wave memory in the case where sources and detector are in an expanding cosmology.  For simplicity, we treat the case where the cosmology is de Sitter spacetime, and discuss the possibility of generalizing our results to the case of a more realistic cosmology.  We find results very similar to those of gravitational wave memory in an asymptotically flat spacetime, but with the magnitude of the effect multiplied  by a redshift factor.

\end{abstract}


\maketitle

\section{Introduction}

Gravitational wave memory, a permanent displacement of the gravitational wave detector after the wave has passed, has been known since the work of Zel'dovich and Polnarev \cite{zeldovich}, extended to the full nonlinear theory of general relativity by Christodoulou \cite{christodoulou}, and treated by several authors \cite{braginsky,epstein,turner,will,favata,lydia1,lydia2,nullfluid,flatmemory,tolwal1,winicour,strominger,flanagan}. 
As is usual in the treatment of isolated systems, all these works consider asymptotically flat spacetimes. However, we do not live in an asymptotically flat spacetime, but rather in an expanding universe.  For sources of gravitational waves whose distance from the detector is small compared to the Hubble radius, modeling the system as an asymptotically flat spacetime should be sufficient.  However, some of the most powerful sources of gravitational waves (e.g. the collision of two supermassive black holes following the merger of their two host galaxies) are at cosmological distances where the asymptotically flat treatment is not sufficient. 

In this paper we will treat gravitational wave memory in an expanding universe.  To avoid the complications of the full nonlinear Einstein equations, our treatment will use perturbation theory.  There is a well developed theory of cosmological perturbations (see e.g. the textbook treatment in \cite{weinberg}).  However, this standard cosmological perturbation theory uses metric perturbations, and we have found \cite{flatmemory} that the properties of gravitational memory are made more clear when using a manifestly gauge invariant perturbation theory based on the Weyl tensor.   Cosmological perturbation theory using the Weyl tensor was developed by Hawking \cite{hawking}.  We will use a treatment similar to that of \cite{hawking}, but also, using the conformal flatness of Friedman-Lema\^itre-Robertson-Walker (FLRW) spacetimes, a treatment that draws heavily on the techniques used in \cite{flatmemory}.  

Cosmological perturbations depend on the equation of state of the matter.  The universe, both at the current time and at any previous times from which a realistic source of gravitational wave memory could come, is dominated by dust and a cosmological constant.  For simplicity, in this treatment we will treat only the case of a cosmological constant, leaving the more general case of dust and a cosmological constant for subsequent work.  Thus this work treats gravitational waves in an expanding de Sitter spacetime.  The perturbation equations are developed in section II, the cosmological memory effect is calculated in section III, and the implications of the results are discussed in section IV.    

\section{equations of motion}
From the Bianchi identity ${\nabla _{[\epsilon }}{R_{\alpha \beta ] \gamma \delta}} =0$
we have
\begin{equation}
{g^{\epsilon \alpha}}{\nabla _\epsilon}{C_{\alpha \beta \gamma \delta}} = {\nabla _{[\gamma}}{S_{\delta ]\beta}}
\label{bianchi}
\end{equation}
where ${S_{\alpha \beta}} = {R_{\alpha \beta}} - {\frac 1 6}R{g_{\alpha \beta}}$ and 
$R = {g^{\alpha \beta}}{R_{\alpha \beta}}$.  Using the Einstein field equation with cosmological constant
\begin{equation}
{R_{\alpha \beta }} - {\textstyle {\frac 1 2}} R {g_{\alpha \beta}} + \Lambda {g_{\alpha \beta}} 
= 8 \pi {T_{\alpha \beta }}
\end{equation}
we find that eqn. (\ref{bianchi}) becomes
\begin{equation}
{g^{\epsilon \alpha}}{\nabla _\epsilon}{C_{\alpha \beta \gamma \delta}} = 8 \pi {\nabla _{[\gamma}}{X_{\delta ]\beta}}
\label{bianchi2}
\end{equation} 
Where ${X_{\alpha \beta}} = {T_{\alpha \beta}} - {\frac 1 3} T {g_{\alpha \beta}}$ and 
$T = {g^{\alpha \beta}}{T_{\alpha \beta}}$.

Both the Weyl tensor, and $T_{\alpha \beta}$ vanish in de Sitter spacetime.  It then follows that 
when we perturb eqn. (\ref{bianchi2}) from a de Sitter background, the perturbed equation takes the same form with the Weyl tensor and stress-energy replaced by their (gauge invariant) perturbations and the metric and derivative operator replaced with their background values.  We will rewrite this perturbed equation in a convenient form making use of the conformal flatness of de Sitter spacetime.  Recall that the line element in a spatially flat Friedman-Lema\^itre-Robertson-Walker (FLRW) spacetime takes the form
\begin{equation}
d{s^2} = - d {t^2} + {a^2}(t) ( d {x^2} + d {y^2} + d {z^2} ) \; \; \; .
\label{flrw1}
\end{equation}    
Then introducing the usual conformal time $\eta$ by $\eta \equiv \int dt/a$, we find that the line element takes the form
\begin{equation}
d{s^2} = {a^2} \left [ - d {\eta^2} +  d {x^2} + d {y^2} + d {z^2} \right ] \; \; \; .
\end{equation}
That is, the de Sitter metric takes the form ${g_{\alpha \beta}} = {a^2} {\eta _{\alpha \beta}}$ where $\eta _{\alpha \beta}$ is the Minkowski metric with Cartesian coordinates $(\eta,x,y,z)$.  It then follows that the perturbed eqn. (\ref{bianchi2}) takes the form
\begin{equation}
{\partial ^\alpha} \left ( {a^{-1}} {C_{\alpha \beta \gamma \delta }} \right ) = 8 \pi \left [ 
a {\partial _{[\gamma}}{X_{\delta ] \beta}} + {X_{\beta [ \gamma}}{\partial _{\delta ]}}a + 
{\eta _{\beta [ \gamma}} {X_{\delta ] \lambda }} {\partial ^\lambda}a \right ] \; \; \; .
\label{bianchi3}
\end{equation}
Here $\partial _\alpha$ is the coordinate derivative operator with respect to the Cartesian coordinates $(\eta,x,y,z)$.  Also here and in what follows we use the convention that indicies are raised and lowered with the Minkowski metric 
$\eta _{\alpha \beta}$.

Following the method of \cite{flatmemory} we now 
decompose all quantities in terms of spatial tensors as follows, using latin letters for spatial indicies: 
\begin{eqnarray}
{E_{ab}} \equiv {a^{-1}}{C_{a\eta b\eta}}
\label{edef}
\\
{B_{ab}} \equiv ({a^{-1}}){\textstyle {\frac 1 2}} {{\epsilon ^{ef}}_a}{C_{efb\eta}}
\label{bdef}
\\
\mu = {T_{\eta \eta}}
\\
{q_a}= {T_{\eta a}}
\\
{U_{ab}} = {T_{ab}}
\end{eqnarray}
Here ${\epsilon _{abc}} = {\epsilon _{\eta abc}}$ where $\epsilon_{\alpha \beta \gamma \delta}$ is the Minkowski spacetime volume element.  
Then eqn. (\ref{bianchi3}) yields two constraint equations
\begin{eqnarray}
{\partial ^b}{E_{ab}} = 4 \pi a ( {\textstyle {\frac 1 3}}{\partial_a}( 2\mu + {{U^c}_c}) - {\partial _\eta} {q_a})
\label{cnstrE}
\\
{\partial ^b}{B_{ab}} = 4 \pi a {{\epsilon ^{ef}}_a} {\partial _e}{q_f}
\label{cnstrB}
\end{eqnarray}
and two equations of motion
\begin{eqnarray}
{\partial _\eta}  {E_{ab}}   - {\textstyle {1 \over 2}} {{\epsilon _a}^{cd}}{\partial_c}{B_{db}} 
- {\textstyle {1 \over 2}} {{\epsilon _b}^{cd}}{\partial_c}{B_{da}} 
\nonumber
\\
= 4 \pi a
\left [ {\partial_{(a}}{q_{b)}} - {\textstyle {\frac 1 3}} {\delta _{ab}}{\partial _c}{q^c}
 - {\partial _\eta} ( {U_{ab}} - {\textstyle {\frac 1 3}} {\delta _{ab}} {{U^c}_c} ) \right ] 
\; + \; 4 \pi {a'} ( {U_{ab}} - {\textstyle {\frac 1 3}} {\delta _{ab}} {{U^c}_c} )
\label{evolveE}
\\
{\partial _\eta} {B_{ab}} +  {\textstyle {1 \over 2}} {{\epsilon _a}^{cd}}{\partial_c}{E_{db}} 
+ {\textstyle {1 \over 2}} {{\epsilon _b}^{cd}}{\partial_c}{E_{da}} 
= 2 \pi a \left (
{{\epsilon _a}^{cd}}{\partial_c}{U_{db}} +  
{{\epsilon _b}^{cd}}{\partial_c}{U_{da}} \right )
\label{evolveB}
\end{eqnarray}
Here ${a'} = da/d\eta$ and $\delta _{ab}$, the Kronecker delta, is the spatial metric of Minkowski spacetime.  

We now want to decompose the spatial tensors into tensors on the two-sphere.  We introduce the usual spherical polar coordinates $(r,\theta,\phi)$ with the usual relation to the Cartesian coordinates $(x,y,z)$.  We use capital latin letters to denote two-sphere components.  From the electric part of the Weyl tensor 
$E_{ab}$ we obtain a scalar $E_{rr}$ as well as a vector and a symmetric, trace-free tensor given by
\begin{eqnarray}
{X_A} = {E_{Ar}}
\\
{{\tilde E}_{AB}} = {E_{AB}} - {\textstyle {1 \over 2}} {H_{AB}} {{E_C}^C}
\end{eqnarray}
Here $H_{AB}$ is the metric on the unit two-sphere, and all two-sphere indicies are raised and lowered with this metric.
Similarly, the decomposition of the magnetic part of the Weyl tensor yields $B_{rr}$ and 
\begin{eqnarray}
{Y_A} = {B_{Ar}}
\\
{{\tilde B}_{AB}} = {B_{AB}} - {\textstyle {1 \over 2}} {H_{AB}} {{B_C}^C}
\end{eqnarray}
The decomposition of the spatial vector $q_a$ yields a two-sphere scalar $q_r$ and vector $q_A$, while the decomposition
of the spatial tensor $U_{ab}$ yields two-sphere scalars $U_{rr}$ and $N \equiv {{U^c}_c}$, 
vector ${V_A} \equiv {U_{Ar}}$ and 
a symmetric trace-free tensor 
\begin{equation}
{W_{AB}} = {U_{AB}} - {\textstyle {1 \over 2}} {H_{AB}} {{U_C}^C}
\end{equation}  
Then the constraint equations (eqns. (\ref{cnstrE}) and (\ref{cnstrB})) become
\begin{eqnarray}
{\partial _r}{E_{rr}} + 3 {r^{-1}}{E_{rr}} + {r^{-2}} {D^A}{X_A} 
= 4 \pi a
\left ( {\textstyle {\frac 1 3}}{\partial _r} ( 2\mu + N) - {\partial _\eta}{q_r} \right  )
\label{cnstra}
\\
{\partial _r}{B_{rr}} + 3 {r^{-1}}{B_{rr}} + {r^{-2}} {D^A}{Y_A} = 4 \pi a
{r^{-2}}{\epsilon ^{AB}}{D_A}{q_B}
\label{cnstrb}
\\
{\partial _r}{X_A} + 2 {r^{-1}} {X_A} - {\textstyle {1 \over 2}} {D_A}{E_{rr}} + {r^{-2}} {D^B}{{\tilde E}_{AB}}
\nonumber
\\
= 4 \pi a ({\textstyle {\frac 1 3}} {D_A} ( 2\mu + N) - {\partial _\eta} {q_A} )
\label{cnstrc}
\\
{\partial _r}{Y_A} + 2 {r^{-1}} {Y_A} - {\textstyle {1 \over 2}} {D_A}{B_{rr}} + {r^{-2}} {D^B}{{\tilde B}_{AB}}
= 4 \pi a {{\epsilon _A}^B}({D_B} {q_r} - {\partial _r} {q_B})
\label{cnstrd}
\end{eqnarray}
Here $D_A$ is the derivative operator and $\epsilon_{AB}$ is the volume element of the unit two-sphere.  
The evolution equations (eqns. (\ref{evolveE}) and (\ref{evolveB})) become
\begin{eqnarray}
{\partial _\eta} {B_{rr}} + {r^{-2}} {\epsilon^{AB}}{D_A}{X_B} 
= 4 \pi a {r^{-2}} {\epsilon^{AB}}{D_A}{V_B}
\label{evolvea}
\\
{\partial _\eta} {E_{rr}}  - {r^{-2}} {\epsilon^{AB}}{D_A}{Y_B} 
\nonumber
\\
=  4 \pi a \left ( {\partial _r}{q_r} - {\partial _\eta}{U_{rr}} + {\textstyle {\frac 1 3}} {\partial _\eta} (N-\mu) \right ) + 4 \pi {a'} \left ( {U_{rr}} - {\textstyle {\frac 1 3}} (2N+\mu) \right )
\label{eovlveb}
\\
{\partial _\eta}{Y_A} +  {\textstyle {1 \over 2}} {r^{-2}} {\epsilon^{CD}}{D_C}{{\tilde E}_{DA}} 
+ {\textstyle {1 \over 4}} {{\epsilon_A}^C} (3 {D_C}{E_{rr}} - 2 {\partial _r}{X_C})
\nonumber
\\
= 2 \pi a \left ({{\epsilon_A}^C} ({\textstyle {\frac 1 2}}{D_C}(3{U_{rr}}-N)-{\partial _r}{V_C})
+ {r^{-2}}{\epsilon ^{BC}}{D_B}{W_{CA}}\right )
\label{evolvec}
\\
{\partial _\eta}{X_A}  - {\textstyle {1 \over 2}} {r^{-2}} {\epsilon^{CD}}{D_C}{{\tilde B}_{DA}} 
- {\textstyle {1 \over 4}}  {{\epsilon_A}^C}(3 {D_C}{B_{rr}} - 2 {\partial _r}{Y_C})
\nonumber
\\
= 2 \pi a ({D_A}{q_r}+{\partial _r}{q_A}) - 4 \pi a ( {r^{-1}}{q_A} + {\partial _\eta}{V_A}) + 4 \pi {a'} {V_A}
\label{evolved}
\\
{\partial _\eta} {{\tilde B}_{AB}} + {\textstyle {1 \over 2}}  
{{\epsilon_A}^C} ({D_C}{X_B}+ {r^{-1}}{{\tilde E}_{CB}} - {\partial _r} {{\tilde E}_{CB}}) 
\nonumber
\\
+ {\textstyle {1 \over 2}}  {{\epsilon_B}^C} 
({D_C}{X_A} + {r^{-1}}{{\tilde E}_{CA}} - {\partial _r} {{\tilde E}_{CA}}) 
+ {\textstyle {1 \over 2}} {H_{AB}} {\epsilon^{CD}}
{D_C}{X_D}
\nonumber
\\
= 2 \pi a {{\epsilon_A}^C}({D_C}{V_B} + {r^{-1}}{W_{CB}}- {\partial _r}{W_{CB}})
\nonumber
\\
+ 2 \pi a {{\epsilon_B}^C}({D_C}{V_A} + {r^{-1}}{W_{CA}}- {\partial _r}{W_{CA}})
+ 2 \pi a {H_{AB}}{\epsilon^{CD}}{D_C}{V_D}
\label{evolvee}
\\
{\partial _\eta}  {{\tilde E}_{AB}}  - {\textstyle {1 \over 2}}  
{{\epsilon_A}^C} ({D_C}{Y_B}- {\partial _r} {{\tilde B}_{CB}}+{r^{-1}} {{\tilde B}_{CB}}) 
\nonumber
\\
- {\textstyle {1 \over 2}}  {{\epsilon_B}^C}  
({D_C}{Y_A}- {\partial _r} {{\tilde B}_{CA}}+{r^{-1}} {{\tilde B}_{CA}}) 
- {\textstyle {1 \over 2}}  {H_{AB}}{\epsilon^{CD}}{D_C}{Y_D}
\nonumber
\\
= 4 \pi a \left (  {D_{(A}}{q_{B)}} - {\textstyle {1 \over 2}} {H_{AB}} {D_C}{q^C}) - {\partial _\eta}{W_{AB}} \right )
+ 4 \pi {a'} {W_{AB}}
\label{evolvef}
\end{eqnarray}

\section{calculation of memory}

We now consider the behavior of the fields at large distances from the source.  Unlike the asymptotically flat case, we cannot make use of the formal definition of null infinity: de Sitter conformal infinity is spacelike, and all gravitational radiation is negligible there.  Instead we define the optical scalar
$u=\eta -r$ and consider the case of large $r$ and moderate values of $u$.  Note that in this case ``large $r$'' means large compared to the wavelength of the gravitational waves emitted by the source, but {\emph not} large compared to the Hubble length.  That is, we treat the case where $r {a'}$ is of order 1.  In the case of Minkowski spacetime, it is shown in 
\cite{flatmemory} that stress-energy gets to large $r$ and moderate $u$ by traveling in null directions: that is, the dominant component of the stress-energy takes the form
\begin{equation}
{T_{\alpha \beta}} = A {\partial _\alpha} u {\partial _\beta} u
\label{nullstress1} 
\end{equation}
We will assume that eqn. (\ref{nullstress1}) also holds in our case.  From conservation of stress-energy, it then follows that $A$ takes the form $ A = L {a^{-2}} {r^{-2}}$ where $L$ is a function of $u$ and the
two-sphere coordinates.  In physical terms, the quantity $L$ is the power radiated per unit solid angle. That is we have
\begin{equation}
\mu =  N =  {U_{rr}} = - {q_r} = L {a^{-2}} {r^{-2}} + \dots
\label{nullstress2}
\end{equation}
with all other components of the stress-energy falling off more rapidly.
Here $\dots$ means ``terms higher order in $r^{-1}$'' 
We assume that, as in the asymptotically flat case, the electric and magnetic parts of the Weyl tensor behave as follows: 
\begin{eqnarray}
{{\tilde E}_{AB}} = {e_{AB}} r + \dots
\\
{{\tilde B}_{AB}} = {b_{AB}}  r + \dots
\\
{X_A} = {x_A}  {r^{-1}} + \dots
\\
{Y_A} = {y_A}  {r^{-1}} + \dots
\\
{E_{rr}} = P  {r^{-3}} + \dots
\\
{B_{rr}} = Q  {r^{-3}} + \dots
\end{eqnarray}
Here the coefficient tensor fields are functions of $u$ and the two-sphere coordinates.  
Furthermore, we assume that, as in the asymptotically flat case, in the limit as $|u| \to \infty$ the only one of these coefficient tensor fields that 
does not vanish is $P$. Note that because
of the relation between Cartesian and spherical coordinates ${\tilde E}_{AB}$ behaving like $r$ corresponds
to Cartesian components of the electric part of the Weyl tensor behaving like $r^{-1}$.  

Now keeping only the dominant terms in eqns. (\ref{cnstra}-\ref{cnstrd}) and using the fact that $r {a'}$ is of order unity, we obtain
\begin{eqnarray}
- {\dot P} + {D^A}{x_A} = - 8 \pi L {a^{-2}} (a+ r {a'})
\label{scricna}
\\
- {\dot Q} + {D^A}{y_A} = 0
\label{scricnb}
\\
- {{\dot x}_A} + {D^B}{e_{AB}} = 0
\label{scricnc}
\\
- {{\dot y}_A} + {D^B}{b_{AB}} = 0
\label{scricnd}
\end{eqnarray}
Here an overdot means derivative with respect to $u$.  Similarly, 
keeping only the dominant terms in eqns. (\ref{evolvea}-\ref{evolvef}) yields
\begin{eqnarray}
{\dot Q} + {\epsilon^{AB}}{D_A}{x_B} = 0 
\label{scrieva}
\\
{\dot P} - {\epsilon^{AB}}{D_A}{y_B} = 8 \pi L {a^{-2}} (a+ r {a'})
\label{scrievb}
\\
{{\dot y}_A} + {\textstyle {1 \over 2}} {\epsilon^{CD}}{D_C}{e_{DA}} +  {\textstyle {1 \over 2}} 
{{\epsilon_A}^C}{{\dot x}_C} = 0
\label{scrievc}
\\
{{\dot x}_A} - {\textstyle {1 \over 2}}{\epsilon^{CD}} {D_C}{b_{DA}} -  {\textstyle {1 \over 2}} 
{{\epsilon_A}^C}{{\dot y}_C} = 0
\label{scrievd}
\\
{{\dot b}_{AB}} + {{\epsilon_A}^C}{{\dot e}_{CB}} = 0
\label{scrieve}
\\
{{\dot e}_{AB}} - {{\epsilon_A}^C}{{\dot b}_{CB}} = 0
\label{scrievf}
\end{eqnarray}

From here, the analysis proceeds essentially as in \cite{flatmemory}.  By convention, the scale factor $a$ is unity at the present time.  Therefore at the position of the detector, $\eta$ is the same as the usual time, $E_{ab}$ (despite the factor of $a^{-1}$ in eqn. (\ref{edef})) is equal to the physical electric part of the Weyl tensor and thus is directly related to tidal force, and $a' $ is equal to $H_0$, the Hubble constant. 
Note that eqn. (\ref{scrievf}) is redundant, since it is equivalent to eqn. (\ref{scrieve}).  
Since $e_{AB}$ and $b_{AB}$ vanish as
$u \to - \infty$, it follows from eqn. (\ref{scrieve}) that
${b_{AB}} = - {{\epsilon_A}^C}{e_{CB}}$.  This can be used to 
eliminate $b_{AB}$ from eqns. (\ref{scricnd}) and (\ref{scrievd}) which then become
\begin{eqnarray}
{{\dot y}_A} +  {\epsilon^{CD}}{D_C}{e_{DA}} = 0
\label{reduce1a}
\\
{{\dot x}_A} - {\textstyle {1 \over 2}} {D^C}{e_{CA}} -  {\textstyle {1 \over 2}} 
{{\epsilon_A}^C}{{\dot y}_C} = 0
\label{reduce1b}
\end{eqnarray}
Combining eqn. (\ref{reduce1a}) with eqn. (\ref{scrievc}) then yields
\begin{equation}
{{\dot y}_A} + {{\epsilon_A}^B}{{\dot x}_B} = 0
\label{reduce2}
\end{equation}
However, since $x_A$ and $y_A$ vanish as $u \to - \infty$,  it then follows from eqn. (\ref{reduce2}) that  
\begin{equation}
{y_A} = - {{\epsilon_A}^B}{x_B}
\end{equation}
Thus, we can eliminate $y_A$ from eqns. (\ref{scricnb}) and (\ref{scrievb}) which then become
\begin{eqnarray}
{\dot Q} + {\epsilon^{AB}}{D_A}{x_B} = 0 
\\
{\dot P} - {D^A}{x_A} =  8 \pi L (1 + r {H_0})
\end{eqnarray}
But these equations are then redundant, since they are equivalent to 
eqns. (\ref{scrieva}) and (\ref{scricna}) respectively.  Thus the only 
independent quantities are ${e_{AB}}, \, {x_A}, \, P, \, Q$ and $L$.  These quantities satisfy the following equations
\begin{eqnarray}
{D^B}{e_{AB}} = {{\dot x}_A}
\label{reduce3a}
\\
{\epsilon^{BC}}{D_B}{e_{CA}} = {{\epsilon_A}^C}{{\dot x}_C}
\label{reduce3b}
\\
{D_A}{x^A} = {\dot P} - 8 \pi L (1 + r {H_0})
\label{reduce3c}
\\
{\epsilon^{AB}}{D_A}{x_B} = - {\dot Q}
\label{reduce3d}
\end{eqnarray}

Now let's consider how to use eqns. (\ref{reduce3a}-\ref{reduce3d}) to find the memory.  
Recall that $e_{AB}$ is (up to a factor involving the distance
and the initial separation) the second time derivative of the separation of the masses.  Thus we want to integrate $e_{AB}$
twice with respect to $u$.  Define the velocity tensor $v_{AB}$, memory tensor $m_{AB}$ and a tensor $z_A$ by
\begin{eqnarray}
{v_{AB}} \equiv {\int _{- \infty} ^u} {e_{AB}} du
\\
{m_{AB}} \equiv {\int _{- \infty} ^\infty} {v_{AB}} du
\\
{z_A} \equiv {\int _{- \infty} ^\infty} {x_A} du
\end{eqnarray} 
Now consider two masses in free fall whose initial separation is $d$ in the $B$ direction.  Then after the wave has passed
they will have an additional separation.  Call the component of that additional separation in the $A$ direction $\Delta d$.
Then it follows from the geodesic deviation equation that
\begin{equation} \label{mem*2}
\Delta d = - {\frac d r} {{m^A}_B}
\end{equation}
To find $m_{AB}$ we first integrate eqns. (\ref{reduce3a}) and (\ref{reduce3b}) to obtain
\begin{eqnarray}
{D^B}{v_{AB}} = {x_A}
\\
{\epsilon^{BC}}{D_B}{v_{CA}} = {{\epsilon_A}^C}{x_C}
\end{eqnarray}
Then integrating again from $-\infty$ to $\infty$ we obtain
\begin{eqnarray}
{D^B}{m_{AB}} = {z_A}
\label{system1a}
\\
{\epsilon^{BC}}{D_B}{m_{CA}} = {{\epsilon_A}^C}{z_C}
\label{system1b}
\end{eqnarray}
Now integrating eqns. (\ref{reduce3c}) and (\ref{reduce3d}) from $-\infty$ to $\infty$ yields
\begin{eqnarray}
{D_A}{z^A} = \Delta P - 8 \pi F (1 + r {H_0})
\label{system1c}
\\
{\epsilon^{AB}}{D_A}{z_B} = 0
\label{system1d}
\end{eqnarray}
where the quantities $\Delta P$ and $F$ are defined by
$\Delta P = P(\infty) - P(-\infty)$ and 
$F = {\int _{-\infty} ^\infty } L du$.  In physical terms, $F$ is the amount of energy radiated per unit solid angle.
In deriving eqn. (\ref{system1d}) we have used the fact that $Q$ vanishes in the limit as $|u| \to \infty$.  Since $z_A$ is curl-free, there must be a scalar $\Phi$ such that 
${z_A} = {D_A} \Phi$.  Then using eqns. (\ref{system1c}) and (\ref{system1a}) we find
\begin{eqnarray}
{D_A}{D^A}\Phi =  \Delta P - 8 \pi  F (1 + r {H_0})
\label{system2a}
\\
{D^B}{m_{AB}} = {D_A}\Phi
\label{system2b}
\end{eqnarray}

\section{Discussion}

We now consider the physical implications of these results, and in particular 
of eqns. (\ref{system2a}-\ref{system2b}).  As in the asymptotically flat case, there are two kinds of gravitational wave memory: an ordinary memory due to sources that do not get out to infinity and a null memory due to sources that do get out to infinity.  The ordinary memory is sourced by $\Delta P$, that is the change in the radial component of the electric part of the Weyl tensor.  The null memory is sourced by $F$, the energy per unit solid angle radiated to infinity.  However, in contrast to the asymptotically flat case, there is a factor of $1+r{H_0}$ multiplying $F$.  Note that in cosmology, the wavelength of light from distant sources is redshifted by a factor of $1+z$ and that to first order in $z$ we have $1+z=1+r{H_0}$.  Thus, expressed in terms of $F$ and $r$ it seems that the null memory is enhanced by a factor of $1+z$.  However, $r$ is not a directly observed property of a distant object: instead we observe the luminosity of the object and infer a luminosity distance $d_L$ related to $r$ by ${d_L}=r(1+z)r$.  Since the observed memory is given by eqn. (\ref{mem*2}) which has a factor of $r^{-1}$, it follows that when expressed in terms of $F$ and $d_L$, the null memory is enhanced by a factor of ${(1+z)}^2$.  However, $F$ is the energy radiated per unit solid angle as measured by the observer, who is at cosmological distance from the source.  Instead, one might want to calculate the local $F$ (which we will call $F_{\rm loc}$) as measured by an obsever who is sufficiently far from the source to be in its wave zone but still at a distance small compared to the Hubble radius.  That is, 
$F_{\rm loc}$ is the $F$ that source would have in Minkowski spacetime.  Because the energy of radiation is diminished by a factor of $1/(1+z)$, it follows that $F={F_{\rm loc}}/(1+z)$.  Therefore, when expressed in terms of $F_{\rm loc}$
and the luminosity distance, the memory is enhanced by a factor of $1+z$.

Finally, we consider the possible generalization of our memory calculation to the case where the background cosmology contains dust in addition to the cosmological constant.  What makes eqn. (\ref{bianchi2}) so simple in a de Sitter background is that the terms on the right hand side involve derivatives of the stress-energy, and that these derivatives vanish in de Sitter spacetime.  However, as shown in \cite{hawking}, for fluid matter the source terms in the equation of motion for the Weyl tensor involve the shear of the fluid four-velocity, and the equation for the Weyl tensor must be supplemented with equations for the shear.  Thus one obtains a set of coupled equations involving the Weyl tensor and the shear.  Nonetheless, it is argued by Thorne \cite{thorne} that for gravitational waves from realistic sources propogating in a cosmological spacetime, one can apply the geometric optics approximation in which the wavelength of the gravitational waves is small compared to all other length scales in the problem.  The implication of this argument for the calculation of gravitational wave memory in the cosmological setting is that one expects all the extra terms in the equations due to shear to be negligible.  This question is currently under study.

\section*{Acknowledgments}

We would like to thank Bob Wald and Abhay Ashtekar for helpful discussions.   
LB is supported by NSF grant DMS-1253149 to The University of Michigan. DG is supported by NSF grants PHY-1205202 
and PHY-1505565 to Oakland University. 
STY is supported by NSF grants DMS-0804454, DMS-0937443 and DMS-1306313 to Harvard University.

\end{document}